\documentclass[conference]{IEEEtran}
\IEEEoverridecommandlockouts

\usepackage{cite}
\usepackage{amsmath,amssymb,amsfonts}
\usepackage{graphicx}
\usepackage{textcomp}
\usepackage{xcolor}
\usepackage{caption}
\usepackage{subcaption}
\usepackage{balance} 
\usepackage{graphics}
\usepackage[T1]{fontenc}
\usepackage{txfonts}
\usepackage{mathptmx}
\usepackage{color}
\usepackage{booktabs}
\usepackage{textcomp}
\usepackage{amsmath}
\usepackage{amsfonts}
\usepackage{duckuments}
\usepackage{bm}
\usepackage{listings}
\usepackage{amsmath}
\usepackage{amsfonts}
\usepackage{multirow}
\usepackage{caption}
\usepackage{verbatim}
\usepackage{todonotes}
\usepackage[noend]{algpseudocode}
\usepackage{microtype}        
\usepackage{ccicons}          

\usepackage{pifont}
\usepackage{enumitem}
\usepackage{url}

\newcommand{\code}[1]{\textbf{\texttt{\scriptsize{#1}}}}

\def\BibTeX{{\rm B\kern-.05em{\sc i\kern-.025em b}\kern-.08em
    T\kern-.1667em\lower.7ex\hbox{E}\kern-.125emX}}
\begin{document}

\title{Guided Optimization for Image Processing Pipelines}

\author{\IEEEauthorblockN{Yuka Ikarashi$^1$ , Jonathan Ragan-Kelley$^1$ , Tsukasa Fukusato$^2$ , Jun Kato$^3$ , Takeo Igarashi$^2$ }
\IEEEauthorblockA{MIT CSAIL$^1$, The University of Tokyo$^2$, AIST$^3$}}

\maketitle

\begin{abstract}
Writing high-performance image processing code is challenging and labor-intensive.
The Halide programming language simplifies this task by decoupling high-level algorithms from ``schedules'' which optimize their implementation.
However, even with this abstraction, it is still challenging for Halide programmers to understand complicated scheduling strategies and productively write valid, optimized schedules. To address this, we propose a programming support method called ``guided optimization.''
Guided optimization provides programmers a set of valid optimization options and interactive feedback about their current choices, which enables them to comprehend and efficiently optimize image processing code without the time-consuming trial-and-error process of traditional text editors.
We implemented a proof-of-concept system, Roly-poly, which integrates guided optimization, program visualization, and schedule cost estimation to support the comprehension and development of efficient Halide image processing code.
We conducted a user study with novice Halide programmers and confirmed that Roly-poly and its guided optimization was informative,
increased productivity, and resulted in higher-performing schedules in less time.
\end{abstract}


\section{Introduction}

Performance optimization is challenging even for experts because the number of optimization choices 
they have is enormous. The barrier is even higher for novices, for whom it is difficult to reason about what choices are valid and what their impact might be, and it is essential to help them break down the choice space to a manageable set of valid options.
Performance and program visualization techniques are investigated widely
in both academia and industry to address this issue, such as always-on
visualization of call stacks~\cite{theseus}, and parallel execution~\cite{cilkpride}.
These methods, however, are text-based, and programmers 
need to carefully write syntactically correct code and manually explore a large space of optimization choices to decide an appropriate optimization strategy.

The Halide programming language~\cite{Ragan-Kelley:2012:DAS:2185520.2185528, halide} is a widely used open-source domain-specific language (DSL) for image processing. It is explicitly designed to achieve high-performance while significantly simplifying the coding process.
Halide decouples high-level \textit{algorithms} from their \textit{schedules}. An \textit{algorithm} is a functional definition of \textit{what} to compute (e.g., a box blur algorithm or corner detection algorithm). \textit{Schedules} define \textit{how} to compute a given algorithm, especially issues related to performance optimization, including choices affecting memory locality and parallelism. Compared with general purpose programming languages such as C++ and Python, this separation allows Halide programmers to write the algorithm part first and optimize its performance independently.

\begin{figure}[t]
  \centering
  \includegraphics[width=0.7\linewidth]{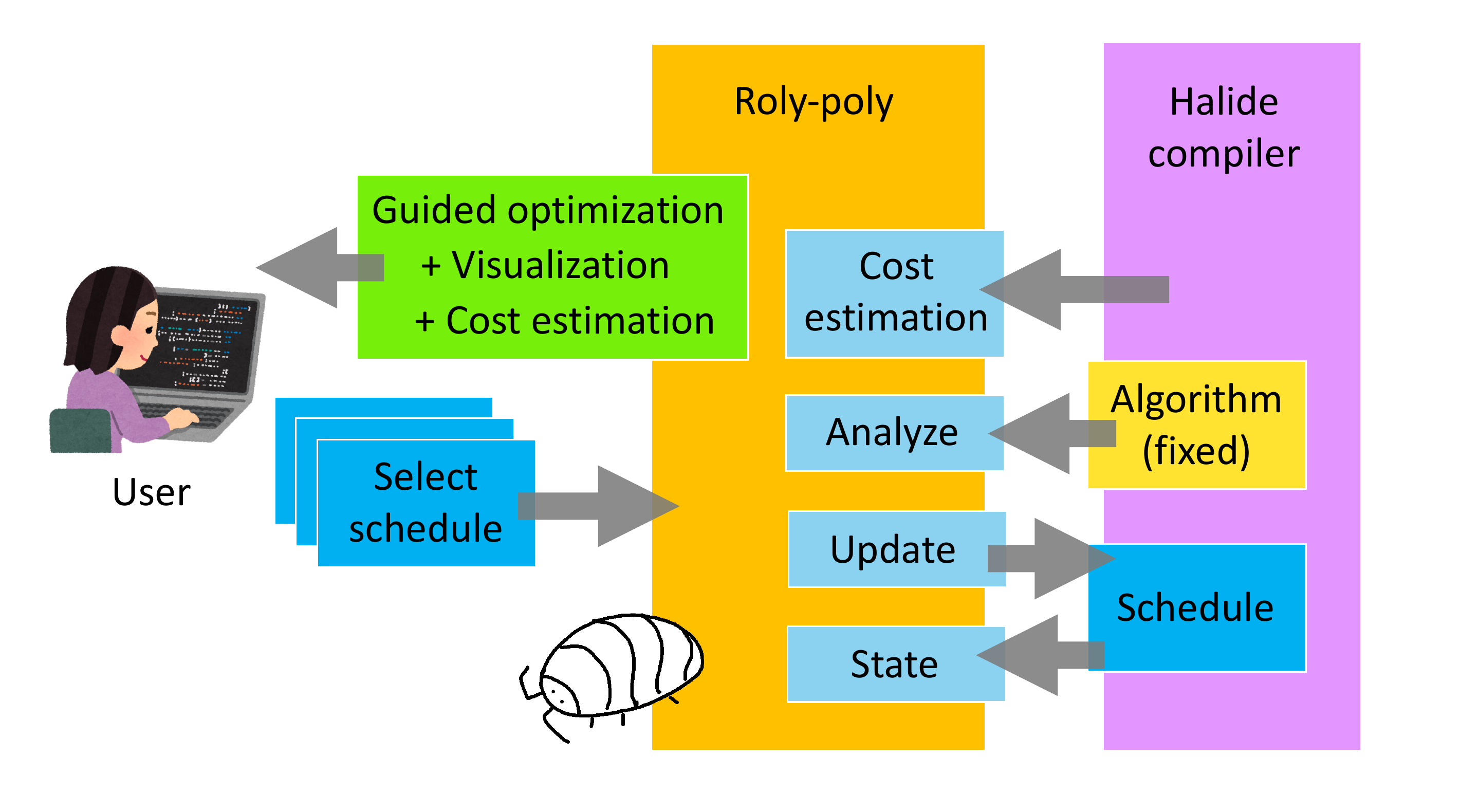}
  \caption{The overview of Roly-poly.}
 \label{fig:roly-poly}
\end{figure}

However, writing optimized schedules in Halide is still challenging.
While the scheduling language is concise, different scheduling choices cause complex, global transformations of the generated code. Understanding what choices are valid, and what effect they will have, is difficult as the implied transformations are invisible to the programmer and their effect must be modeled in the programmer's head.
It is also difficult to predict how schedules will perform. The most efficient schedules require complex trade-offs between parallelism, locality, and the total amount of computation in different places throughout a program.
To find the best result, programmers need to repeatedly change scheduling code, compile and execute the program, benchmark the result, and inspect the generated code.
Ongoing research explores removing this burden by automatically inferring fast schedules with no user involvement~\cite{autoschedule}, but these techniques are still limited, and manual programming of schedules remains the standard practice in industry to retain explicit control.

To make authoring schedules more accessible and productive,
we present an interface and workflow for authoring schedules called \textit{guided optimization}.
Guided optimization presents multiple valid scheduling options to the user, together with interactive feedback on the current schedule state. This allows users to directly explore the space of valid schedules without struggling to mentally model the impact of individual scheduling operations, and to make more effective choices based on integrated feedback about their performance impacts.
With such systematic help, we believe that Halide programmers can optimize image processing code more productively and effectively than with traditional manual scheduling. 
This paper introduces our prototype implementation, Roly-poly, to show the feasibility and effectiveness of the proposed method. Roly-poly also features program visualization and scheduling cost estimation to support guided optimization. 
Figure~\ref{fig:roly-poly} shows the system overview of Roly-poly.

This paper makes the following contributions:
\begin{itemize}
    \item The proposal of \textit{guided optimization}, a semi-automatic scheduling approach where the system enumerates a series of valid scheduling choices and asks users to choose one of the options at each step.
    \item An integration of cost estimation into guided optimization.
    \item A visualization of the scheduling strategy at each step of the guided optimization.
    \item An interview with Halide experts and a user study with Halide novices on the effectiveness of guided optimization.
\end{itemize}


\section{Image Processing Scheduling in Halide}
\label{halide}


Halide is embedded in C++ and can be used to build computational photography and computer vision applications efficiently. It decouples image processing code into an \textit{algorithm}, which is a mathematical representation of image pixel transformation, and into a \textit{schedule}, which dictates the order of computation and storage. The program output depends only on the algorithm, and changing schedules solely affects the performance. This separation allows programmers to explore various schedules after defining an algorithm without unintentionally changing the program output.

\begin{figure}
\begin{subfigure}[b]{0.5\textwidth}
\begin{flushleft}
\code{\\
Var x, y;\\
Func kernel, bounded, blur\_y, blur; \\
kernel(x) = exp($-$x$\ast$x/(2$\ast$sigma$\ast$sigma))/ \\
\hspace*{1cm}(sqrtf(2$\ast$ M\_PI) $\ast$ sigma); \\
bounded = BoundaryConditions::repeat\_edge(input);\\
blur\_y(x, y) = (kernel(0) $\ast$ input(x, y) +\\
\hspace*{1cm}kernel(1) $\ast$ (bounded(x, y-1) + bounded(x, y+1)) +\\
\hspace*{1cm}kernel(2) $\ast$ (bounded(x, y-2) + bounded(x, y+2)) +\\
\hspace*{1cm}kernel(3) $\ast$ (bounded(x, y-3) + bounded(x, y+3))); \\
blur(x, y) = (kernel(0) $\ast$ \hspace{0.8mm}blur\_y(x, y) +\\
\hspace*{1cm}kernel(1) $\ast$ (blur\_y(x-1, y) + blur\_y(x+1, y)) +\\
\hspace*{1cm}kernel(2) $\ast$ (blur\_y(x-2, y) + blur\_y(x+2, y)) +\\
\hspace*{1cm}kernel(3) $\ast$ (blur\_y(x-3, y) + blur\_y(x+3, y))); \\
}
\end{flushleft}
\end{subfigure}
\caption*{Algorithm 1: Gaussian Blur algorithm written in Halide.}
\label{list:gaussian}
\end{figure}

\begin{figure}[t]
  \centering
  \includegraphics[width=0.6\linewidth]{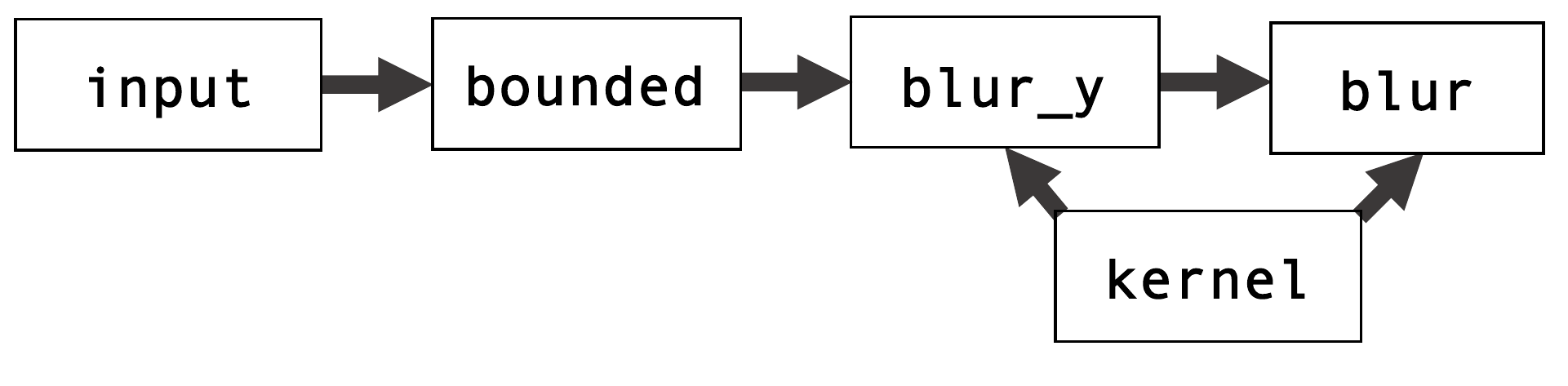}
  \caption{A dependency graph of a Gaussian Blur algorithm.}
  \label{fig:dag}
\end{figure}

\subsection{Scheduling Example with Gaussian Blur}
\label{schedule_gaussian}

We describe basic scheduling concepts by considering an example of a Gaussian Blur algorithm run on a 256x256 image.
Algorithm 1 shows a Gaussian algorithm written in Halide, and Figure~\ref{fig:dag} shows its dependency graph. It first sets a boundary condition on the input image and stores the result in \code{bounded}. Then, it applies a one-dimensional Gaussian kernel function vertically, storing the result in \code{blur\_y}, and then applies a Gaussian kernel horizontally to \code{blur\_y} to produce \code{blur}.

The default schedule generates the following implementation:
\code{\\
\hspace*{ 5mm}for x in 0..255\\
\hspace*{10mm}for y in 0..255\\
\hspace*{15mm}blur $(\cdots)$ = (kernel(0) $\ast$ \hspace{0.8mm}blur\_y(x, y) + $\cdots$;\\
}
All functions are inlined to \code{blur}, and the loop simply iterates \code{x} and \code{y}. This \textit{inlining} schedule has a maximal memory locality because seven elements of \code{blur\_y} are produced and immediately are consumed by \code{blur} (likely readout of the CPU's registers). However, this schedule also induces a maximal redundant computation. This schedule does not cache the result of \code{blur\_y}, and it is computed each time it is used from \code{blur} with the same arguments.

A schedule method called \textit{pre-computation} can be used to reduce redundant computation. For example:
\code{\\
\hspace*{ 5mm}for x in 0..255\\
\hspace*{10mm}for y in 0..255\\
\hspace*{15mm}blur\_y $(\cdots) = \cdots$; \\
\hspace*{ 5mm}for x in 0..255\\
\hspace*{10mm}for y in 0..255\\
\hspace*{15mm}blur $(\cdots) = \cdots$;\\
}
In this schedule, \code{blur\_y} is pre-computed before \code{blur}. This schedule has minimal redundant computations because all required elements of \code{blur\_y} are pre-computed and stored in memory before they are subsequently loaded to compute \code{blur}. However, memory locality is sacrificed because an entire pre-computed image is usually too large to fit in a cache.

Finally, \textit{tiling} can be used to balance the advantages and the disadvantages of \textit{inlining} and \textit{pre-computation}. For example:
\code{\\
\hspace*{ 5mm}for x\_outer in 0..31\\
\hspace*{10mm}for y\_outer in 0..15\\
\hspace*{15mm}for x\_inner in 0..7\\
\hspace*{20mm}for y\_inner in 0..15\\
\hspace*{25mm}blur\_y $(\cdots) = \cdots$;\\
\hspace*{15mm}for x\_inner in 0..7\\
\hspace*{20mm}for y\_inner in 0..15\\
\hspace*{25mm}blur $(\cdots) = \cdots$;\\
}
Tiling divides a loop into an outer-loop and a smaller inner-loop.
In this example, the outer-loop variables of \code{x\_outer} and \code{y\_outer} divide the input image into 512 (32x16) tiles. Each of these tiles has a size of 8x16. In this paper, we call the for-loop range (\code{x\_outer} range of 32 and \code{y\_outer} range of 16) a \textit{tile range} and the divided image size (8x16 in this example) a \textit{tile size}. The tile range is usually just called ``for-loop range,'' but we use this word to distinguish and emphasize that this for-loop exists only for dividing the image into smaller tiles.
In short, if a for-loop of function \code{child} is nested inside a for-loop of function \code{parent}, (a tile size of \code{child}) = (tile size of \code{parent}) / (tile range of \code{child}).

This schedule pre-computes \code{blur\_y} for each tile of \code{blur}, which saves redundant computations while taking advantage of the memory locality of \code{blur\_y}. Memory locality is important for efficient scheduling because loading data from a cache is significantly faster than loading it from the main memory. However, taking a too small tile size will result in redundant computation at the edge of each tiling. Therefore, calibrating the tile size so that it is small enough to fit in a cache and large enough to avoid too many redundant computations is essential for balancing redundant computation and memory locality trade-off.

To summarize, choosing the function's i) compute location in the for-loop (pre-computing or inlining) and ii) tile range are key methods for image processing optimization.

\subsection{Expert Interview}
\label{expert}

To understand the representative workflow of manual scheduling optimization and to collect advice in creating a tool for novice image processing programmers, we conducted interviews with Halide scheduling experts.

We invited three experts aged $25$ to $30$ years with at least two years of Halide experience (one of them had an experience as a teaching assistant 
at an image processing class teaching Halide) and asked them to perform manual scheduling on Unsharp Masking (see Algorithm 2).
The interview was conducted informally, asking them to explain the reasons for their schedule choices.
We chose this task because its scheduling choices were complex enough to measure experts' realistic scheduling procedures.
After the participants completed the task, we asked them how they felt about manual scheduling process and what they felt short in the current programming environment.

On average, an interview took 26 minutes, and the average runtime of a schedule was 32 ms. We learned from the interview the basic strategy of scheduling, which we explained in Section \ref{schedule_gaussian}.
Participants told us that the iterative process of editing, compiling, and executing to check the runtime was laborious and said that they would like to use a tool integrating them into one screen with instant feedback. They also stated that even Halide experts often find it challenging to imagine the CPU and cache state given a schedule, and said that visualization would certainly be useful to picture them.
Based on their comments, we designed an initial concept of Roly-poly.

\section{Related Work}

\subsection{Performance Visualization}
Performance visualization techniques help programmers improve the time and energy
efficiency of their programs~\cite{stateofart}.
Callgrind and Kcachegrind are tools that output the profile information about cache events and the dynamic call graph of the execution~\cite{cachegrind}.
The Linux kernel supports the \textit{perf} profiling tool, which displays CPU usage~\cite{perf}. \textit{gprof} implements more aesthetically pleasing visualization~\cite{gprof}, and Intel's VTune Amplifier is proprietary software used widely in professional high-performance computing (HPC) settings~\cite{vtune}.
These systems focus on low-level visualizations, such as processor and memory visualization. Giving high-level (for-loop level) visualizations and suggestions are beyond the scope of these techniques. We believe that providing users with more abstract-level visual feedback will help them better comprehend image processing optimization.
Moreover, these techniques target general programming languages and environments and do not utilize domain knowledge. Our target domain is image processing, which enables us to provide domain-specific optimization and visualization.

The HPC community suggested many memory visualization techniques~\cite{stateofart} such as visualization of cache hit and misses~\cite{memory1, memory2}, dynamic memory allocation~\cite{vdma}, and the complex relationships between memory performance factors~\cite{memaxes}.
Halide visualization tools highlights computed memory region represented as image cells in each execution step~\cite{hviz}, \cite{traceviz}.
Although these techniques are powerful in visualizing the cache or memory state at one particular moment, their visualizations are low level, and they tend to express temporal transitions by continuous animation or not visualizing them at all. This limits users' ease of understanding because they still need to observe the animation and guess what the performance bottleneck is by manually comparing the source code and the visualization. We envision addressing this issue in Roly-poly by visualizing the whole image processing pipeline as a single static image.

\subsection{Program Visualization in General}
\label{sec:programv}
Program visualization maps programs to visual representations by analyzing source code, executable binaries, and data~\cite{pv}.
Omnicode visualizes the entire history of all runtime values for all program variables~\cite{omnicode}. DejaVu shows an interactive video-player-like interface to visualize the value changes of user-selected variables~\cite{dejavu}. Hoffswell et al. proposed to embed runtime value visualization within the text-based code~\cite{vega}. Projection Boxes allows on-the-fly customization of visualizing runtime values of expressions at the code editor cursor~\cite{projectionboxes}. These research focuses on comprehensible visualization of data structures and variables, and performance visualization is beyond their scope.

\subsection{Program Visualization for Performance Improvement}
Few studies have pushed the boundaries and explored program visualizations aimed at performance improvement.
PerformanceHat~\cite{performancehat} and Beck et al.~\cite{insitu} augments the source code with runtime performance traces. 
Theseus enables always-on program visualization of call stacks and its interactive exploration during the web-based application development~\cite{theseus}.
Projection Boxes shows program visualization next to the code and enables its flexible customization at the programmer's will~\cite{projectionboxes}.
Cilkpride interactively updates the IDE with outputs of the Cilk performance profiler for parallel programs~\cite{cilkpride}.
Skyline is a domain-specific IDE for deep neural network training programs that shows performance predictions and visualizations~\cite{skyline}.
In adjacent research areas such as live coding, live programming, and exploratory programming, performance visualization is not yet actively investigated~\cite{exploratory}.

Roly-poly aims to achieve a similar goal as these works but is different for the following reasons:
1) Existing methods require users to write optimization code in text editors and provide performance visualization as an editor extension. Writing performance optimization code on text editors have an advantage of being highly customizable, but users are not free from making syntactical mistakes and still need to decide an optimization strategy within a large space of optimization choices. Roly-poly's guided optimization controls the image processing optimization workflow itself, and sequentially provides a valid set of scheduling choices to users. The optimization choice a user can make is thus limited to valid schedules, and it is impossible to make syntactic errors in Roly-poly.
2) Roly-poly provides the performance cost estimation of each scheduling option, which is the first attempt in performance visualization literature. The key difference between Roly-poly and existing systems such as Cilkpride and Skyline is that Roly-poly's cost does not show the current schedule's performance, but it shows the predicted cost of the entire resulting pipeline. Thus, we believe that Roly-poly's cost feedback is more global than existing systems.

\subsection{Neural Network Visualization}
In the field of machine learning research, several methods have been proposed to visualize complex neural networks~\cite{Hiddenlayer, PlotNeuralNet, NN-SVG, Tensorspace, tensorflowgraph}. In contrast to traditional performance visualization techniques, most neural network visualization approaches show spatial and temporal information in one image, thus visualizing the whole pipeline. This approach helps users understand the overall system instead of one particular moment. These visualizations can be applied to 2D array computation in general, including our target domain of image processing pipelines. Inspired by these visualizations, Roly-poly takes a similar approach. Instead of animating the visualization, we envision visualizing the whole execution of the image processing pipeline in a single view.

\subsection{Visual Programming Language}
Visual programming languages (VPLs) use an interactive and graphical approach to help users avoid syntactic errors. Agentsheets utilizes a drag-and-drop mechanism to prevent syntactic mistakes~\cite{agentsheets}. Alice~\cite{alice} and Scratch~\cite{scratch} are block-based VPLs for novice programmers to learn the basic principles of programming by combining the provided blocks.
Weintrop and Wilensky found that block-based programming tools were easier to use for high school students than text-based alternatives~\cite{block}.

Inspired by the prior work on VPLs, Roly-poly offers an interactive block-like visualization of the execution of the image processing pipeline.
Compared with manual text-based scheduling, we narrow users' freedom by only providing a set of choices. However, our design choices are that using visual constraints in Roly-poly will help users focus only on essential ideas, not on syntactical mistakes.
Also, some VPLs assign colors to categories based on their functionalities (e.g., control blocks and operators blocks). This inspired Roly-poly to assign distinguished colors to different functions.

\section{Design and Implementation of Roly-Poly}

\begin{figure*}[t]
  \centering
  \includegraphics[width=0.7\linewidth]{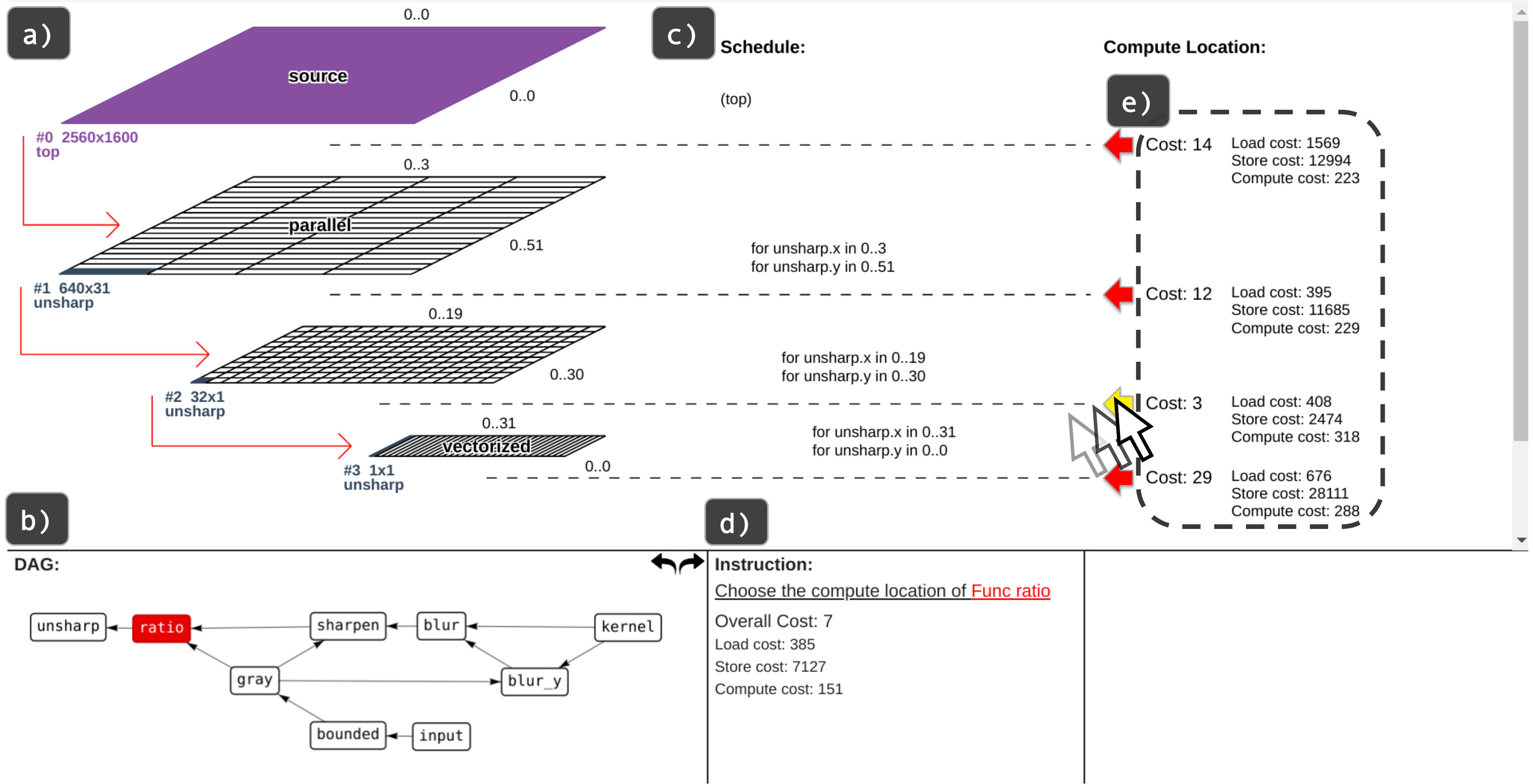}
  \caption{A snapshot of Roly-poly's user interface during the scheduling task of  Unsharp Masking (see Algorithm 2). (a) Visualization of tile sizes. Each tile corresponds to a for-loop block in (c). (b) A dependency graph of the functions obtained by a static analysis of the algorithm. (c) Schedule section showing the for-loop blocks of the current schedule. (d) Instruction from the system. (e) Scheduling cost estimation of possible compute locations.}
  \label{fig:overview}
\end{figure*}
\begin{figure}[t]
  \centering
  \begin{subfigure}[b]{0.29\textwidth}
    \includegraphics[width=\textwidth]{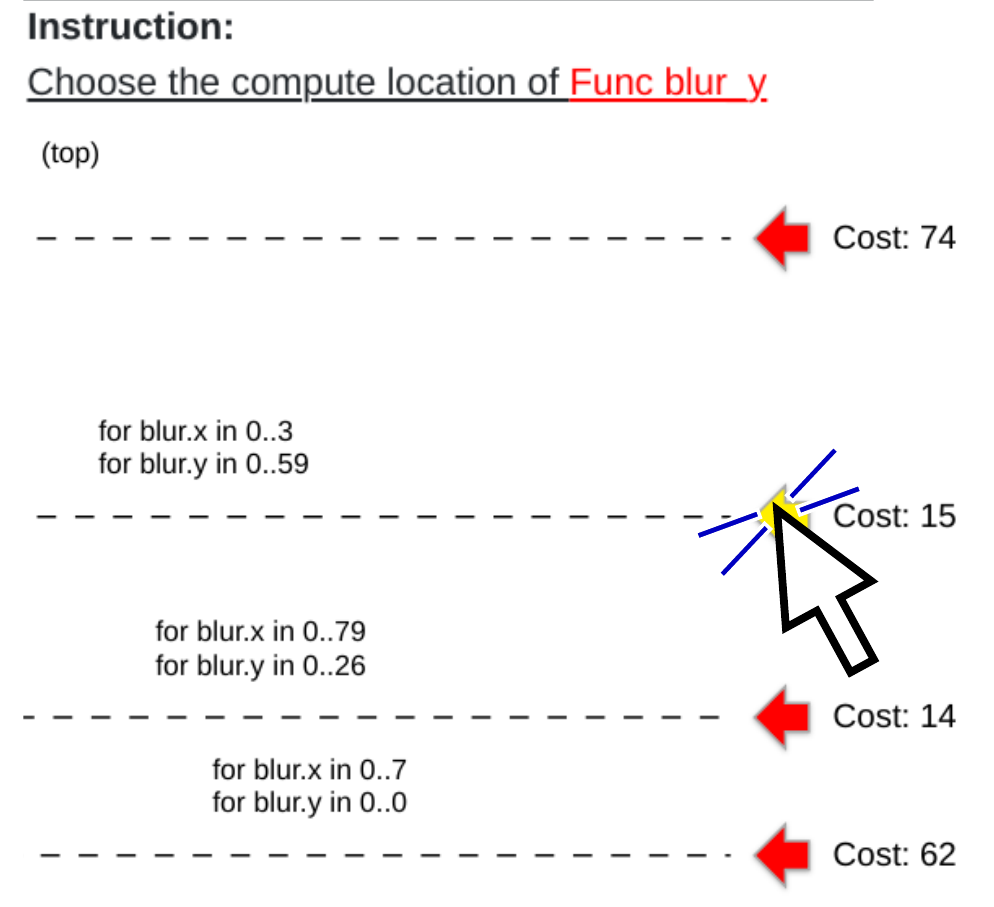}
    \caption{}
    \label{fig:computeloc}
  \end{subfigure}
  \begin{subfigure}[b]{0.25\textwidth}
    \includegraphics[width=\textwidth]{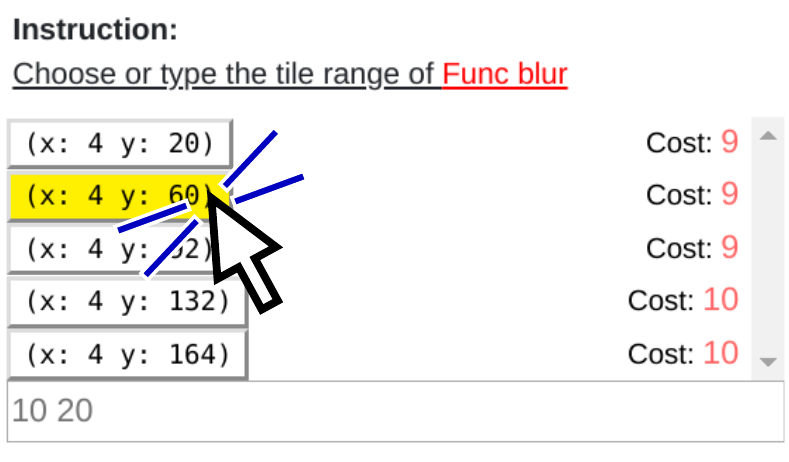}
    \caption{}
    \label{fig:tiling}
  \end{subfigure}
  \begin{subfigure}[b]{0.15\textwidth}
    \includegraphics[width=\textwidth]{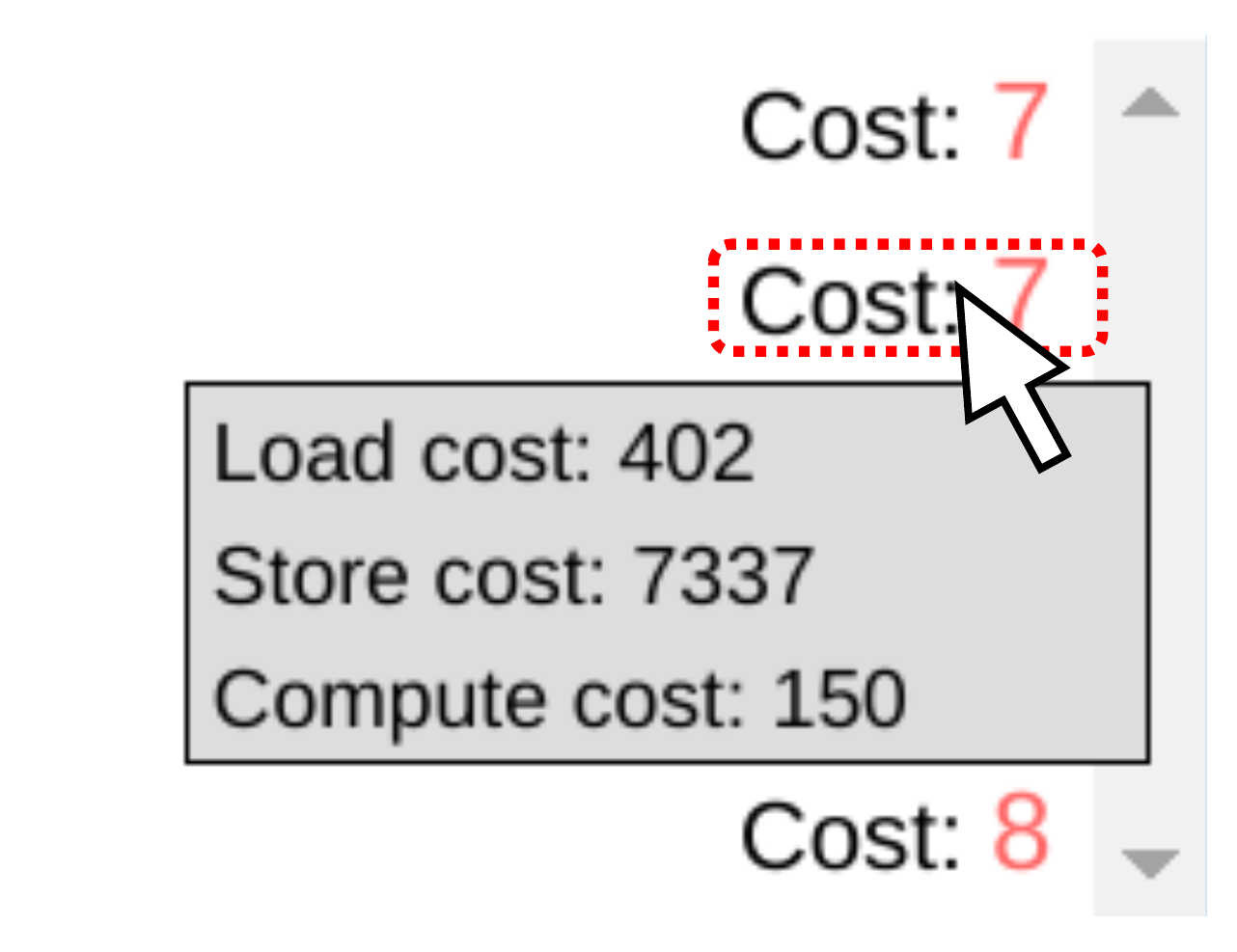}
    \label{fig:cost}
    \caption{}
  \end{subfigure}
    \caption{A snapshot of (a) the compute location selection phase and (b) tile range selection phase  in the Gaussian Blur task. (c) The cost details can be shown in both phases.}
\end{figure}

Our proof-of-concept system, Roly-poly, is divided into the front-end and back-end. The front-end of Roly-poly is implemented as a desktop application built with Electron\footnote{\url{https://www.electronjs.org}}, which communicates with the always-on back-end Halide compiler.
We utilized the Halide compiler's program generation infrastructure for cost estimation and elimination of illegal schedules.
Its program generator supports the essential image processing optimization methods of tiling and pre-computation.

\subsection{System Overview}

Figure~\ref{fig:overview} shows a screenshot of Roly-poly during the scheduling task of Unsharp Masking (see Algorithm 2).
Section (a) shows a visualization of the tile sizes.
Section (b) shows a dependency graph of the functions obtained by a static analysis of the algorithm.
This is an automatically generated graph representing the functional dependencies, which is the same concept as in Figure~\ref{fig:dag}.
It updates the currently scheduled function with a red background (\code{ratio} is highlighted in Figure~\ref{fig:overview} (b)).
This dependency graph visualization enables users understand the algorithm dependencies at a glance, without switching back and forth between the source code and schedule views.

Section (c) shows a for-loop representation of the current schedule.
Each for-loop block corresponds to a tile size visualization in section (a).
Section (d) gives instructions for what users should do next.
In the instructions, the currently scheduled function is highlighted in red (\code{ratio} is highlighted in this case).
Section (e) shows a scheduling cost estimation by the compiler.
It is important to note that these costs are just ``estimates'' and do not always match with the actual performance.
So, the users still need to experiment with various schedules to find optimal schedule, using these cost estimation as a hint.

\subsection{Guided Optimization}

Guided optimization is a semi-automatic method that assists manual scheduling. 
Roly-poly provides step-by-step instructions to a user with a set of valid options.
Users can experiment with various ``valid'' scheduling possibilities just by selecting options presented by the system.

As we explained in the earlier section, the key factors to consider in image processing optimization are \textit{tiling} and \textit{compute location}. There are two scheduling phases in this system to support them. \textit{The compute location selection phase} asks users to choose the function's compute location (see Figure~\ref{fig:computeloc}), and \textit{the tile range selection phase} allows them to choose or type the tile ranges (see Figure~\ref{fig:tiling}).
The system can only be one of these two phases.
Users follow the system's instructions and keep choosing from the suggestions until the scheduling of all functions has been completed.

Users start scheduling an output function (e.g., \code{blur} in Algorithm~1 and \code{unsharp} in Algorithm~2) and continue scheduling functions following an inverse topological order from the output to input (e.g., \code{input} in Algorithm~1 and 2). For each function, users select a compute location and a tile range. When tiling does not apply to a function, the system skips the tiling phase.

\subsubsection{Compute location selection phase}

The schedule section (Figure~\ref{fig:overview} (c,e) and Figure~\ref{fig:computeloc}) shows a for-loop block of the current schedule. In the compute location selection phase, users choose the function's compute location by clicking an arrow on the right side.
If a user clicks \code{Cost 3} in Figure~\ref{fig:computeloc}, the system communicates with the Halide compiler and schedules Func~\code{sharpen} between first and the second inner-most for-loop block in Figure~\ref{fig:computeloc}.
Tile sizes of for-loop blocks in (c) is visualized in (a).
The cost of each compute location is shown on the right side.

\subsubsection{Tile range selection phase}

In the tile range selection phase, users choose a function's tile range, as shown in Figure~\ref{fig:tiling}. Users either click the tile range suggestion or enter the custom tile range in the input box. The estimated cost of each tile range is shown on the right side. The tile range panel is hidden in the compute location selection phase, and it appears in the tile range selection phase.

Each function is initially scheduled to be computed by two-fold for-loops, an external loop and a vectorized inner loop. The external loop is further divided into outer external loop and inner external loop. The tile range selection phase lets the user specify how to divide the external loop into the two loops. The user specifies the tile range of the outer loop and the system automatically computes the tile range of the inner loop.
We made a design choice to let users select the tile \textit{ranges} instead of letting them directly select the tile \textit{sizes}, because it is more intuitive for novice Halide programmers to specify the ranges of for-loop than defining the unfamiliar concept of tile sizes.

The system enumerates multiples of four for width and height to suggest tile ranges. It generates \code{(tile\_size\_width\_of\_a\_parent/4)} $\times$ \code{(tile\_size\_height\_of\_a\_parent/4)} number of tile ranges in total and calculates the estimated cost for each one. Then, the system sorts them in cost ascending order (the lowest cost first) and suggests the top five tile ranges.
We empirically found that users often do not look at suggestions below the fifth one because they prefer entering a custom tile range to scrolling through too many options to explore their schedules.

\subsubsection{Visualization}
As users change the schedule, the visualization panel~(a) is updated to reflect the current schedule. It visualizes the tile size of each for-loop block in the schedule section~(c).
Each function is assigned a color that is consistent throughout the scheduling process and in the visualization (see Figure~\ref{fig:overview}~(a)).

We divide the parent's tile or image size by the children's tile range to obtain the children's tile size.
For example, the image size is 2560$\times$1600 in Figure~\ref{fig:overview}. The range of the for-loop block \#1 is (0$\ldots$3)$\times$(0$\ldots$51), so the tile size of tile \#1 becomes 640$\times$31 (x: 2560/4 = 640, y: 1600/52 $\approx$ 31). Similarly, the tile size of tile \#2 is 32$\times$1  (640/20 = 32, 31/31 = 1).

Tiles with \code{parallel} are automatically parallelized.
Tiles with \code{vectorized} and the corresponding for-loop block are the innermost for-loop and they are automatically vectorized to use SIMD.
Parallelization and vectorization are done by the compiler when applicable and users do not need to specify them.
The system uses the Snap.svg JavaScript SVG library to render each tile.

\subsubsection{Cost Estimation}

Roly-poly uses a cost model generated by backend Halide compiler to interactively estimate the performance characteristics of different choices.
Guided optimization guides users to schedule functions one by one, and the cost of not yet scheduled functions are set default to the cost of inline schedule.
The estimated total cost and cost details are shown in both the compute location selection phase and the tile range selection phase. It also shows the estimated cost of the current schedule below the instruction section (see Figure~\ref{fig:overview}~(d)). The estimated total cost is calculated by many internal parameters, and three main parameters are shown.
The load cost is a cost of loading from memory, the store cost is a cost of storing to the memory, and the compute cost is a cost of CPU computation.

\subsection{Usage Scenario}
We will explain the scheduling flow step-by-step, taking an example of one particular scheduling flow in a Gaussian Blur optimization task (see Algorithm 1). Suppose that a programmer called Sofia is optimizing a Gaussian Blur using Roly-poly.

She first loads an algorithm file into the system. The system analyzes the algorithm, creates a default schedule, and presents an initial user interface. Note that users do not need to manually write any scheduling code before and throughout the scheduling process. The system then gives the first instruction ``Choose or type the tile range of Func blur.'' She notices ``Func blur'' in the instruction and in the dependency graph are highlighted in red. She then understands that she needs to choose the tile range for the highlighted function. In the bottom right of the UI, she sees five buttons with \code{x}, \code{y}, and two numbers. Those buttons are highlighted in yellow when she mouse-overs them, so she notices that these are clickable. She also notices that ``Cost:'' and numbers are shown on the right side of buttons. She thinks that a lower cost indicates a better runtime, but she wants to try a slightly larger tiling size as an experiment, so she chooses the second top tile size with ``Cost: 9.'' The instruction and the tile range suggestion in this stage are shown in Figure~\ref{fig:tiling}.

After clicking the tile range, the system gives a new instruction ``Choose the compute location of Func blur\_y.'' She notices that the highlight of a function dependency graph changed to \code{blur\_y}. She mouse-overs the arrows on the right side of scheduling section, noticing that those are buttons and are clickable. She understands that the tile size of this for-loop block is shown on the left, in the visualization. Then, she explores the system and compares the costs and decides to choose an arrow with cost 15 because she wants to compute \code{blur\_y} in the outer loop.
The actual schedule section in this stage is shown in Figure~\ref{fig:computeloc}.

After she clicks the button, the instruction tells her to ``Choose the compute location of Func bounded.'' She compares the costs and tile sizes in visualization, and chooses the third arrow. After that, the instruction says, ``Choose a compute location of Func kernel.'' She chooses the first arrow because she wants to pre-compute \code{kernel} before the other for-loops. Finally, she chooses Func \code{kernel}'s tile range in response to an instruction ``Choose or type the tile range of Func kernel.'' After that, she notices that the scheduling is completed because the instruction says ``Done!''

\section{User Study}
We conducted a user study with novice Halide programmers to compare the effectiveness of Roly-poly and the manual scheduling operation.

\subsection{Procedure}

\begin{figure}
\begin{subfigure}[b]{0.6\textwidth}
\begin{flushleft}
\code{\\
Var x, y, c;\\
Func kernel, bounded, gray, blur\_y, blur, sharpen, ratio, unsharp;\\
kernel(x) = exp(-x$\ast$x /(2 $\ast$ sigma $\ast$ sigma)) /\\
        \hspace*{1.cm}(sqrtf(2 $\ast$ M\_PI) $\ast$ sigma);\\
bounded = BoundaryConditions::repeat\_edge(input);\\
gray(x, y) = 0.299f $\ast$ bounded(x, y, 0) +\\
                    \hspace*{1cm}0.587f $\ast$ bounded(x, y, 1) +\\
                    \hspace*{1cm}0.114f $\ast$ bounded(x, y, 2);\\
blur\_y(x, y) = (kernel(0) $\ast$ \hspace*{1mm}gray(x, y) +\\
            \hspace*{1cm}kernel(1) $\ast$ (gray(x, y-1) + gray(x, y+1)) +\\
            \hspace*{1cm}kernel(2) $\ast$ (gray(x, y-2) + gray(x, y+2)) +\\
            \hspace*{1cm}kernel(3) $\ast$ (gray(x, y-3) + gray(x, y+3)));\\
blur(x, y) = (kernel(0) $\ast$ \hspace*{0.8mm}blur\_y(x  , y) +\\
                    \hspace*{1cm}kernel(1) $\ast$ (blur\_y(x-1, y) + blur\_y(x+1, y)) +\\
                    \hspace*{1cm}kernel(2) $\ast$ (blur\_y(x-2, y) + blur\_y(x+2, y)) +\\
                    \hspace*{1cm}kernel(3) $\ast$ (blur\_y(x-3, y) + blur\_y(x+3, y)));\\
sharpen(x, y) = 2 $\ast$ gray(x, y) - blur(x, y);\\
ratio(x, y) = sharpen(x, y) / gray(x, y);\\
unsharp(x, y, c) = ratio(x, y) $\ast$ input(x, y, c);\\
}
\end{flushleft}
\end{subfigure}
\caption*{Algorithm 2: Unsharp Masking algorithm written in Halide.}
\label{list:unsharp}
\end{figure}

We asked 10 participants (P1, P2, $\cdots$, P10) to perform the Unsharp Masking scheduling task with Roly-poly and manual scheduling in a within-subjects design.
Participants had eight years of programming experience on average.
All participants had professional programming experience in image processing for commercial and research purposes.
All participants had no previous experience with the Halide programming language and none of them participated in the expert interview (Section~\ref{expert}).

Each participant was invited to a 60-minute session. We conducted all of our studies online using Chrome Remote Desktop.
We first gave 20-minute tutorials on the basic concepts of image processing in Halide. The introduction covered the content of Section~\ref{halide} which we presented earlier, such as \textit{to consider the cache locality} and \textit{to avoid redundant computation.}
We provided the Unsharp Masking algorithm, as shown in Algorithm 2, and asked participants to write schedules and optimize its runtime as much as possible within 15 minutes. For each task, half of the participants were asked to perform the scheduling with Roly-poly first, and the other half were asked to perform scheduling manually first.
In the manual scheduling task, we provided an example code of two methods (\code{compute\_at(...)} for pre-computing a function and \code{tile(...)} for tiling a function) and allowed them to browse the Halide tutorial if they wanted. After the scheduling session, we spent 10 minutes giving them a questionnaire and a semi-structured interview. We recorded their manipulation process and analyzed it afterward.

\begin{figure}[t]
  \centering
  \includegraphics[width=0.77\linewidth]{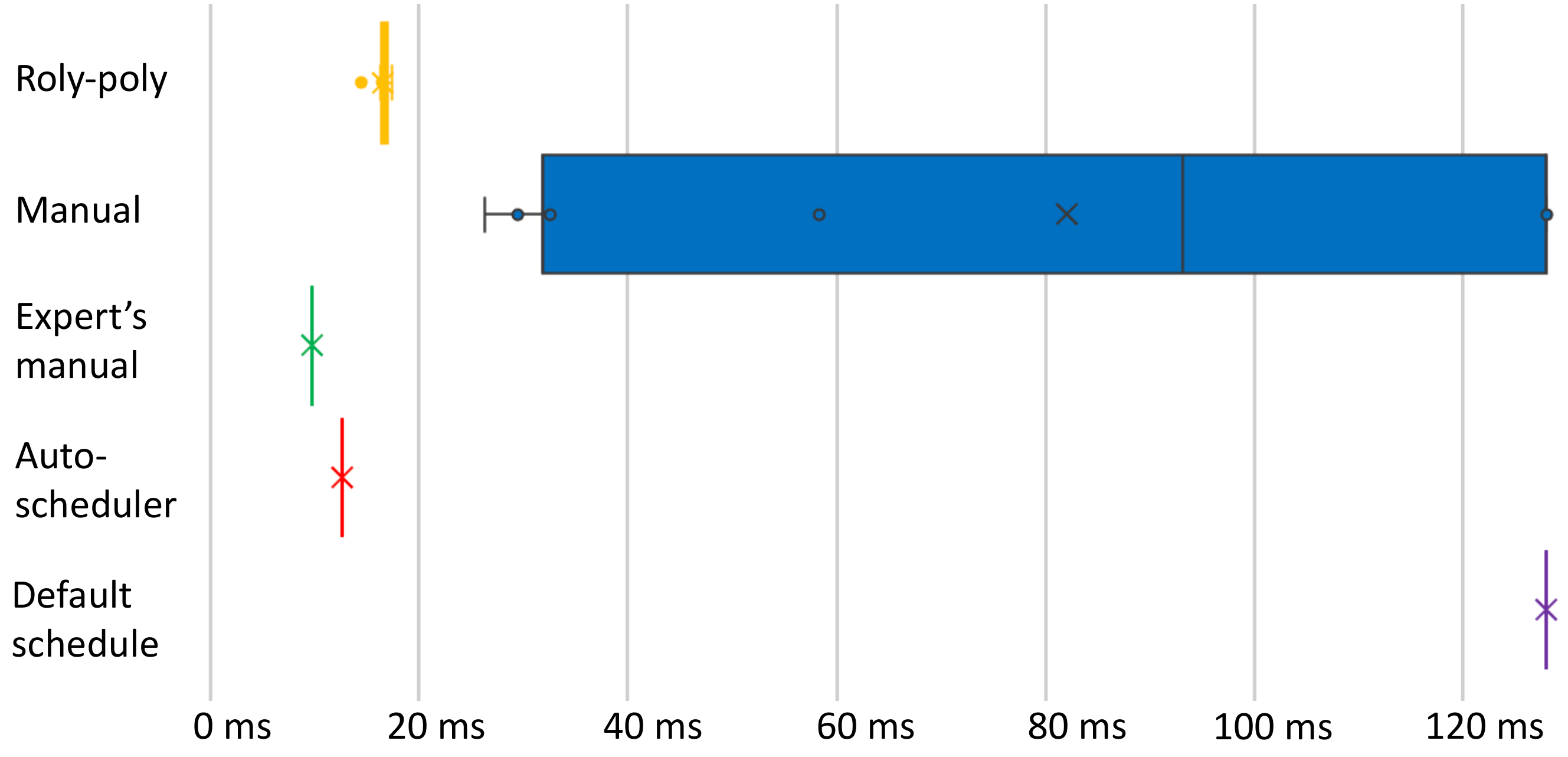}
  \caption{A box and whisker plot showing the distribution of the best runtime of unsharp masking by participants with manual and Roly-poly scheduling, the best runtime from expert interview (see Section~\ref{expert}), runtime from the auto-scheduler, and runtime of the default inline schedule.}
  \label{fig:us-result}
\end{figure}

\begin{figure}[t]
  \centering
  \includegraphics[width=0.77\linewidth]{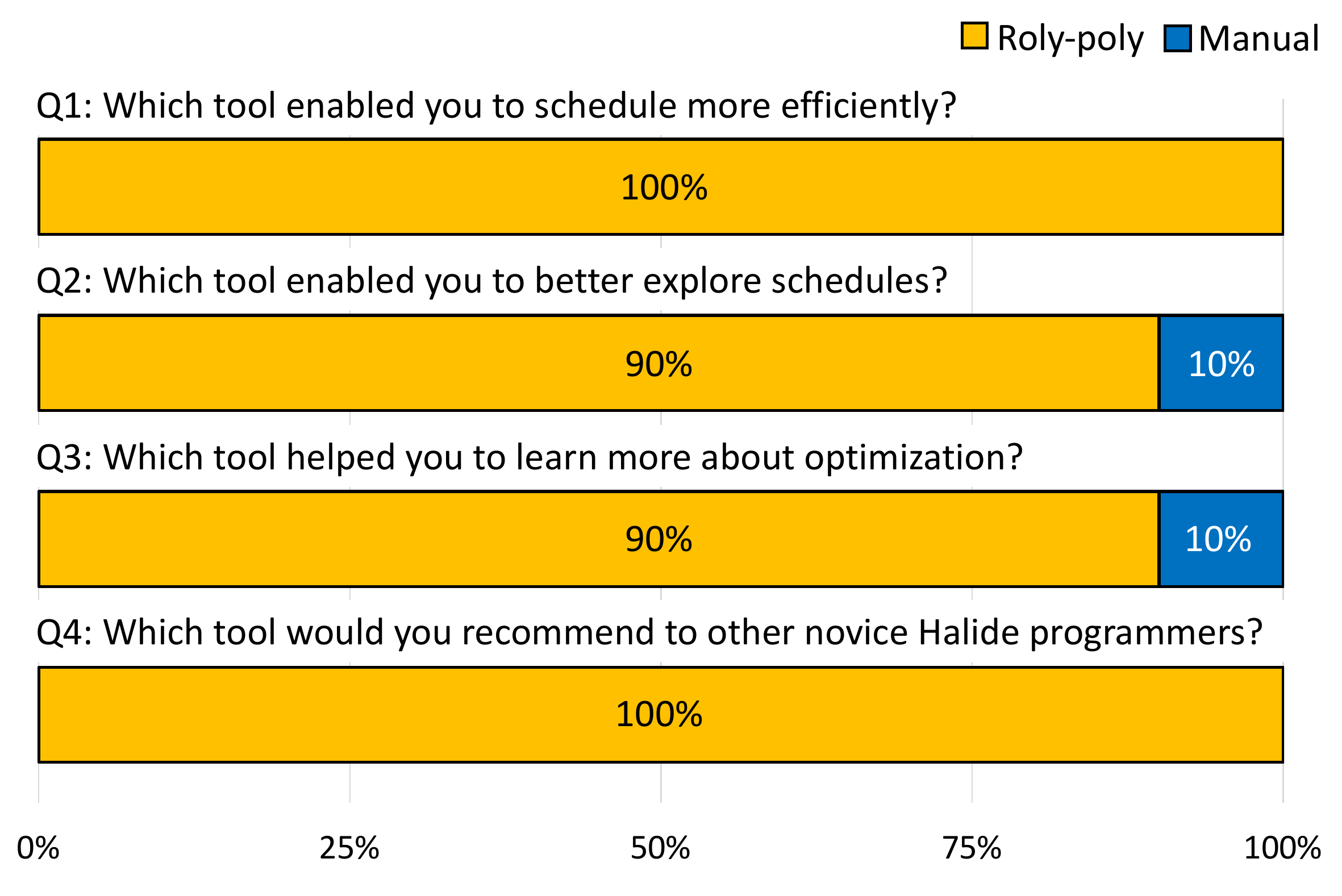}
  \caption{The results of user study questionnaire.}
  \label{fig:us-questions}
\end{figure}

\subsection{Overview of Results}
Figure~\ref{fig:us-result} is a box and whisker plot showing the distribution of the best runtime of unsharp masking by participants with manual and Roly-poly scheduling, the best runtime from expert interview (Section~\ref{expert}), runtime from the auto-scheduler~\cite{autoschedule}, and runtime of the default inline schedule.
Halide provides the default inlining schedule for algorithms without any user-specified schedule, and the runtime of the Unsharp Masking default schedule was 128 ms.
From the user study, five participants could not write any faster code than the default schedule with manual, and all participants achieved better runtime with Roly-poly than with manual.
The average runtime of the unsharp masking with Roly-poly was 16.5 ms, and the average runtime with manual was 82 ms. The best runtime from expert interview was 9.8ms and runtime from the auto-scheduler was 12.7 ms.
Thus the result was better in the order of expert, auto-scheduler, Roly-poly, manual, and default.
This result supports our initial motivation that the expert's manually scheduled runtime is faster than auto-scheduler's.
Moreover, while novices were not quite able to beat the auto-scheduler in their first 15 minutes with Roly-poly, they already come surprisingly close, while their results with manual scheduling are several times slower.

We analyzed the recording video to understand the challenges faced by participants on manual scheduling.
We found that six participants spent most of their time becoming familiar with Halide scheduling syntax, such as how each schedule function take different arguments. Four participants quickly understood the syntax and spent their time figuring out good schedules by trying different schedules.

Figure~\ref{fig:us-questions} shows the post-study questionnaire results. All participants answered that they were able to schedule more efficiently with Roly-poly. Nine participants answered that they could better explore schedules with Roly-poly. Nine participants answered that they could learn more about optimization with Roly-poly. All participants answered that they would recommend Roly-poly to other novice Halide programmers over the manual scheduling. 

Below are the results of semi-structured interviews.
All participants stated that they found manual scheduling challenging, reasoning that the search space was too large, or they felt the effect of each method was a black box. P5, P7, and P9 gave overall comments on their experience.

\begin{itemize}
    \item P9: \textit{I prefer this tool (Roly-poly) more as a Halide scheduling beginner, as it does a lot of hand-holding for people who haven't done scheduling before. The concept and effect of the scheduling methods (e.g., tile size, compute location) were clearly visualized and provided a more intuitive explanation to learn about scheduling.}
    \item P5: \textit{I was not familiar with typical code structures in image processing optimization nor Halide syntax, and Roly-Poly as a visual programming tool helped me in overcoming these obstacles.}
    \item P7: \textit{It was difficult to understand how the manual scheduling works because the search space was too large, and I couldn't find out which function I should have tiled or set compute location to. In the end, it kind of got faster by trying out some methods, but I didn't understand at all how it got faster.}
\end{itemize}


\subsection{Guided Optimization}

All participants said that the guided optimization features were especially useful in Roly-poly.
In particular, P9 and P6 said that the features of sequentially scheduling functions and the instruction telling the next step were useful.
\begin{itemize}
    \item P9: \textit{Guided optimization gave me a clear idea of which function should be scheduled in what order and prevents a lot of potential errors from happening.}
    \item P6: \textit{Guided optimization helped me build a step-by-step strategy for optimization. Necessary information were displayed at the right time of the optimization step.}
\end{itemize}

P7 and P10 said that the guided optimization feature of selecting a schedule from valid options was useful because they did not have to understand the Halide syntax and the details of image processing optimization.
\begin{itemize}
    \item P7: \textit{For novice users like myself, limiting the search space to valid choices was more understandable.}
    \item P10: \textit{Roly-poly was a better starting point for image processing optimization as I didn't need to fully understand the Halide language syntax. The optimization process was self-contained within a GUI.}
\end{itemize}

P1 and P6 said that Roly-poly was useful because it gave them faster and interactive feedback, which enabled them to explore more schedules.
\begin{itemize}
    \item P1: \textit{Roly-Poly gave me faster and more interactive feedback, so I had more time to do trial-and-error within the time limit, which helped me better understand the scheduling as a result.}
    \item P6: \textit{I had to go back and forth between the tutorial and code with manual scheduling, which took me a while to experiment. Roly-poly made it easy to do what I wanted to do and especially made it easier to explore.}
\end{itemize}

However, on a more negative note, P3 mentioned that it was ``too kind'' and that it might prevent users from learning. Also, P5 mentioned that if he was not a novice he would prefer using the manual scheduling, even though he found that Roly-poly was useful as a beginner.
\begin{itemize}
    \item P3: \textit{When we consider teaching or learning the optimization in Halide, I think Roly-Poly is `too kind' for users (annoying things, which is important for programming in many cases, are almost completely hidden!). For example, I encountered compile errors when I tried manual scheduling. However, Roly-Poly hides such errors (it doesn't allow me to choose invalid compute location), and this would prevent users from learning things such as `Why is this compute location invalid?'}
    \item P5: \textit{Roly-Poly enables a more friendly way to explore arguments for me, but just in a subspace of the scheduling space. If I am not a novice, maybe I will prefer to use manual scheduling.}
\end{itemize}

\subsection{Cost Estimation}

Eight participants said that the cost estimation helped them select schedules from valid options, because it helped them narrow down the optimization space and made it easier to understand the relationship between scheduling methods and the resulting optimization.
\begin{itemize}
    \item P1: \textit{The cost estimation was really informative and helpful to make decisions on scheduling. I could understand which choice I should make at each step.}
    \item P7: \textit{Roly-poly showed costs in detail, showing total costs, memory costs, and compute costs. I could prioritize regarding each details.}
\end{itemize}

\noindent
P9, however, added that the cost could be deceiving.
\begin{itemize}
    \item P9: \textit{Cost estimation was useful but could be deceiving. It provides you with `probably good' guesses that basically avoid irrational choices, but it also discourage the user from exploring other options which could lead to a potentially better solution.}
\end{itemize}
This comment provided the insight that the cost suggestion might prevent users from exploring the scheduling space freely and deepening their understanding of scheduling strategies when they just follow the best cost estimation.

\subsection{Visualization}

Participants found that the visualization of tile sizes was useful, because it gave a meaning to for-loop blocks. P3 and P9 mentioned that the visualization helped them comprehend the one to one correspondence of for-loop blocks and tile sizes. P7 mentioned that the visualization made the compute location selection phase clearer.
\begin{itemize}
    \item P3: \textit{I understood from the visualization that the input image was split into multiple tiles and each tile is also split into multiple sub-tiles. The visualization also told me how functions were executed by step-by-step execution procedure.}
    \item P9: \textit{Associating the visualization result (tile sizes) with the for-loop block and showing them in the same row gave me an intuitive idea on how each decision affected the scheduling process. }
    \item P7: \textit{It was clear from the arrow and the dashed line that I can click the arrow and execute the current Func between for-loop blocks.}
\end{itemize}

\noindent
However, P6 found that the visualization was hard to grasp at a glance.
\begin{itemize}
    \item P6: \textit{Visualization was helpful when I examined closely, but when the scheduling pipeline became more complex, it displayed too much information on the screen and it was slightly difficult to keep track of each tile size.}
\end{itemize}


\section{Discussion and Future Work}

We learned from the user study that cost suggestions might lead users to blindly follow the best cost without understanding why. It is necessary to assess this effect by conducting a user study with and without the cost suggestion and identify the best way to present a cost to foster users' understanding. For example, showing the cost at first and hiding it after several attempts might help users' understanding.


We utilized the Halide compiler's program generator to generate valid schedule options. This approach has the advantage that we can show the estimated cost, but it has the disadvantage of limiting the user's search space to that of program generator's. Although the key optimization methods of tiling and pre-computation are supported, advanced methods such as fusing and cloning are not supported by the program generator. Moreover, Roly-poly guides users through the functions sequentially, which forces users to schedule functions in a given order. We plan to investigate whether it is useful to give users the freedom to choose the order of functions to schedule, or whether it is too much detail.

Roly-poly's guided optimization and the visualization can be ported to other algorithm-schedule decoupled languages such as Taichi~\cite{taichi} and Tiramisu~\cite{tiramisu} by writing schedule generators for them.
One area of future work that Roly-poly has inspired is to create a guided optimization system for these languages and also for different application domains, such as optimization for machine learning.

\section{Conclusion}
We presented an interactive scheduling system that features guided optimization as well as cost estimation and program visualization.
Guided optimization lets users select tile ranges and compute locations from valid choices step-by-step and provides interactive feedback of the current schedule,
thereby helping Halide programmers achieve good performance.
We conducted a user study to assess the effectiveness and limitations of Roly-poly by recruiting novice Halide programmers. All participants preferred Roly-poly to manual when it came to optimization efficiency and learning, and many of them appreciated the guided optimization features. Cost estimation received mixed feedback, and we suggested some possible research directions for further investigation.
We hope that Roly-poly encourages further development of interactive scheduling systems in the future.

\section*{Acknowledgments}
This work was supported by JSPS KAKENHI Grant Number JP19K20316 and JST CREST Grant Number JPMJCR20D4, Japan.

\end{document}